\documentclass[twocolumn,secnumarabic,amssymb, nobibnotes,aps, prd]{revtex4-1}
\usepackage[utf8]{inputenc}

\usepackage{mathtools}
\usepackage{amsmath}
\usepackage[version=4]{mhchem}
\usepackage{color}
\usepackage{appendix}
\usepackage{bm}
\usepackage{units}
\usepackage{upgreek}

\newcommand{\Eqref}[1]{Eq.~\ref{#1}}

\newcommand{\figref}[1]{Fig.~\ref{#1}}
\newcommand{\RNum}[1]{\textup{\uppercase\expandafter{\romannumeral #1\relax}}}
\newcommand{\mo}[1]{\ensuremath{w_{#1}}}
\newcommand{\ins}{\mathrm{in}}
\newcommand{\outs}{\mathrm{out}}
\newcommand{\Vs}{V_\mathrm{sys}}
\newcommand{\Vd}{V}
\newcommand{\kBT}{k_\mathrm{B}T}

\newcommand{\sF}{\ensuremath{s^\mathrm{f}}}
\newcommand{\sB}{\ensuremath{s^\mathrm{b}}}

\begin{document}
	
	\title{Controlling biomolecular condensates via chemical reactions}
	
	\author{Jan Kirschbaum, David Zwicker}%
	\affiliation{Max Planck Institute for Dynamics and Self-Organization}
	\date{\today}%
	
	\begin{abstract}
		Biomolecular condensates are small droplets forming spontaneously in biological cells via phase separation. They play a role in many cellular processes, but it is unclear how cells control them. Cellular regulation often relies on post-translational modifications of proteins. For biomolecular condensates, such chemical modifications could alter the molecular interaction of key condensate components. We here test this idea using a theoretical model based on non-equilibrium thermodynamics. In particular, we describe the chemical reactions using transition-state theory, which accounts for the non-ideality of phase separation. We identify that fast control, like in cell signaling, is only possible when external energy input drives the reaction out of equilibrium. If this reaction differs inside and outside the droplet, it is even possible to control droplet sizes. Such an imbalance in the reaction could be created by enzymes localizing to the droplet. Since this situation is typical inside cells, we speculate that our proposed mechanism is used to stabilize multiple droplets with independently controlled size and count. Our model provides a novel and thermodynamically consistent framework for describing droplets subject to non-equilibrium chemical reactions.
	\end{abstract}
	
	\maketitle
	
	\section*{Introduction}
	
	Biomolecular condensates are small droplets that structure the cell interior of eukaryotes ~\cite{Brangwynne_Science_2009,berry_RepProgPhys_2018} and prokaryotes~\cite{greening_natrev_2020,cohan_trendsbio_2020,azaldegui_biophysJ_2020}.
	They form by phase separation and participate in a wide range of cellular functions~\cite{alberti_JourCellSci_2017}:
	Since they are chemically distinct from their surrounding, they can act as reaction centers~\cite{lyon_natrev_2020,Peeples_biorxiv_2020}, like the nucleolus inside the nucleus~\cite{lafontaine_natrevmolcel_2020}.
	In particular, locally elevated concentrations can induce polymerization, like in microtubule branching~\cite{setru_nature_2021} or in centrosomes~\cite{Zwicker_pnas_2014,woodruff_cell_2017} that additionally control the subcellular organization.
	Condensates can also store molecules to buffer fluctuations in gene expression~\cite{klosin_science_2020} or to release them later when the condensate dissolves, like germ granules and the Balbiani body~\cite{schuh_TrCellBio_2021}.
	Condensates also help to detect changes in the environment externally, like receptor clusters~\cite{zhao_DevCell_2020,jaqaman_CurrOpCellBio_2021}, and internally, like stress granules~\cite{alberti_JMolBio_2018}.
	In particular, transcriptional condensates actively regulate gene expression~\cite{sabari_DevCell_2020}.
	In all these examples, the cell controls the size, position, or count of the biomolecular condensates~\cite{berry_RepProgPhys_2018,snead_MolCell_2019}.
	The formation of biomolecular condensates can be described in the framework of liquid-liquid phase separation~\cite{hyman_annrefcellbio_2014}.
	This implies that the droplet size is determined by the total amount of droplet material.
	Moreover, inevitable surface tension driven Ostwald ripening, which is a coarsening process dissolving smaller droplets in favor of larger ones, so that only a single droplet remains in thermodynamic equilibrium.
	The theory can also be used to predict how the droplet size depends on global parameters, like temperature, pH, and salt concentration~\cite{Choi_AnnRevBiophys_2020,adame_bpj_2020,Brangwynne_Nature_2015}.
	Cells can directly control condensates by changing protein concentrations or molecular interactions~\cite{riback_nature_2020, dignon_annrevphyschem_2020}.
	The interactions are mainly dictated by the genetic sequence, which varies on evolutionary time scales.
	On cellular time scales, post-translational modifications can further adjust the interactions, enabling more dynamic regulation~\cite{hofweber_friend_2019,hondele_emtoplife_2020}.
	As an example, phosphorylating the carboxy-terminal domain (CTD) of RNA polymerase II dissolves CTD droplets in vitro~\cite{Boehning_nature_2018}.
	More generally, chemical reactions, like such post-translational modifications, can affect the dynamics of droplets and explain how cells could regulate condensate size, location and count~\cite{rai_kinase-controlled_2018,Soeding_TCB_2020}.
	
	Theoretical studies of active droplets, which combine phase separation and chemical reactions, suggest that chemical reactions can suppress Ostwald ripening, leading to coexisting droplets of similar size~\cite{Zwicker_PRE_2015,Wurtz_PRL_2018} and even droplet division~\cite{Zwicker_nature_2016}.
	These studies described chemical reactions using fixed rate laws, which does not include the molecular interactions necessary for phase separation.
	Instead, a thermodynamically consistent theory is necessary to faithfully describe the interplay of phase separation with reactions.
	Earlier work in this direction~\cite{Weber_RepProgPhys_2019,Carati_1997,bazant_faraday_2017} suggests that reactions need to be driven away from equilibrium to be effective.
	
	We here present a minimal model of active droplets, which combines non-equilibrium thermodynamics~\cite{DeGroot_NEQTD_1984,julicher_repprogphys_2018} and transition state theory~\cite{Pagonabarraga_PhysA_1997,hanggi_revmodphys_1990} to describe the chemical reactions.
	It focuses on chemical potentials as key quantities and describes the non-equilibrium driving explicitly.
	We identify the conditions under which droplet size control is possible and determine the associated energetic cost.
	In the following, we build up the complete model by starting from passive liquid-liquid phase separation and we then successively add the reaction, the driving, and enzymatic control.

	\section*{Modelling phase separation with chemical transitions}
	
	\begin{figure}
		\centering
		\includegraphics[width=\linewidth]{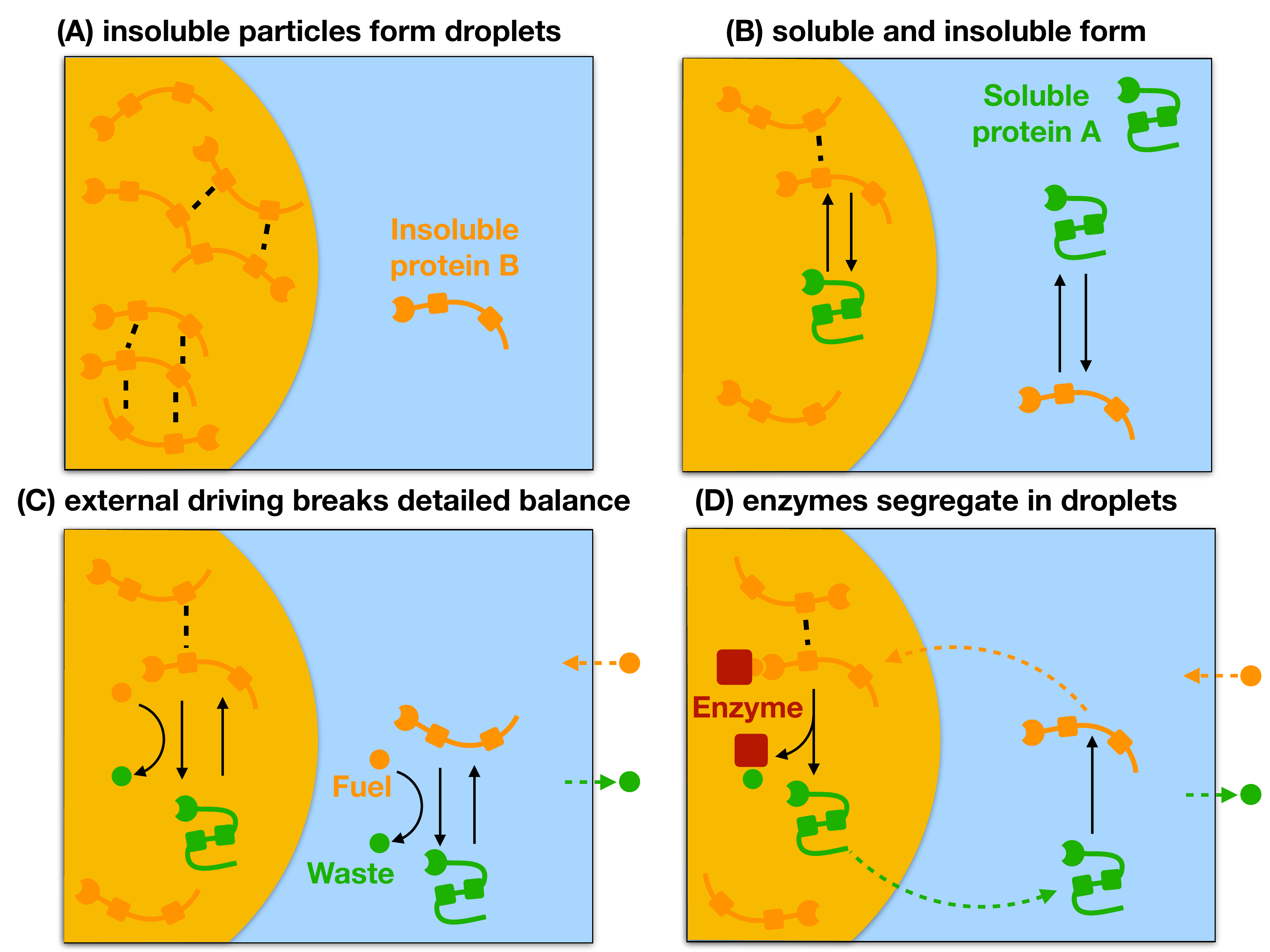}
		\caption{
			\textbf{Schematic representation of the four discussed models:}
			\textbf{(A)} $B$ molecules (orange) with weak enthalpic interactions (black dashed lines) form droplets (orange region) that coexist with the dilute phase (blue).
			\textbf{(B)} A spontaneous chemical transition (black arrow) between the segregating form $B$ and the soluble form $A$ (green) determines the available amount of droplet material $B$.
			\textbf{(C)} A second reaction reaction (curved arrow) driven by the conversion of fuel $F$ (orange circle) to waste $W$ (green circle) can lead to a non-equilibrium stationary state when $F$ and $W$ are coupled to particle baths (arrows across box).
			\textbf{(D)} An enzyme segregating into droplets (red square) controlling the driven reaction causes cyclic diffusive fluxes (dashed arrows) in the system, which can stabilize multiple droplets in the same system.
		}
		\label{fig:A0}
	\end{figure}
	
	We consider an incompressible, liquid mixture of a solvent and a chemical component that can exist in two different forms: a form $A$, which is soluble in the solvent, and an insoluble form $B$, which segregates from the solvent.
	The composition of the system is then given by the volume fractions $\phi_i(\bm{x})$ of components $i=A,B$ at each position $\bm{x}$.
	They evolve as
	\begin{subequations}
		\label{Eq:Contin}
		\begin{align}
		\partial_t \phi_A &= -\nabla \bm{j}_A - s\\
		\partial_t \phi_B &= -\nabla \bm{j}_B + s\;,
		\end{align}
	\end{subequations}
	where $\bm{j}_i$ are diffusive fluxes and $s$ is the reactive flux associated with the chemical transition $A \rightleftharpoons B$.
	We assume that the chemical component cannot leave the system, which implies the normal fluxes $\bm n.\bm j_A$ and $\bm n .\bm j_B$ vanish at the boundary with normal vector $\bm n$.
	Consequently, the total amount of the chemical component is conserved.

	The diffusive and reactive fluxes, $\bm j_i$ and $s$, can be described in the framework of non-equilibrium thermodynamics~\cite{DeGroot_NEQTD_1984}, which ensures that a closed system relaxes to thermodynamic equilibrium and that detailed balance is obeyed.
	One consequence is that the fluxes $\bm{j}_i$ and $s$ are related to the chemical potentials $\mu_i(\bm{x})$ of the species $i=A,B$.
	In particular, the diffusive fluxes can be approximated by 
	$\bm{j}_i = -\sum_j\Lambda_{ij}\nabla \mu_j$, where the diffusive mobilities $\Lambda_{ij}$ form the symmetric, positive semi-definite Onsager matrix~\cite{DeGroot_NEQTD_1984}.
	In contrast, such a linear approximation is inadequate for the reactive flux~$s$ \cite{DeGroot_NEQTD_1984} and we thus discuss a more detailed model below.
	
	The chemical potentials $\mu_i(\bm{x})$ describe how the free energy $F$ of the system changes when a particle~$i=A,B$ replaces an equal volume of solvent at position $\bm{x}$.
	They are thus given by $\mu_i = v_i \, \delta F[\phi_A, \phi_B] / \delta \phi_i$, where we consider constant molecular volumes $v_i$  and we assume $v_A=v_B$.
	Since the molecular interactions are typically short-ranged, the free energy of this isothermal system of temperature~$T$ can be expressed as
	\begin{align}\label{Eq:F}
	F[\phi_A, \phi_B] = \int \biggl[
	f(\phi_A, \phi_B)
	-\!\!\!\!\sum_{i,j=A,B,C}\frac{\kappa_{ij}}{2}\nabla\phi_i . \nabla\phi_j
	\biggr] \mathrm{d}V \;,
	\end{align}
	where the integral is over the entire system of volume~$\Vs$.
	Here, $f$ is the local free energy density, which governs phase separation~\cite{Weber_RepProgPhys_2019}.
	Conversely, $\kappa_{ij}$ penalizes composition gradients, which results in surface tension effects~\cite{Mao_SoftMatter_2019}.
	As a concrete example, we consider the free energy density
	\begin{align}
	\frac{f(\phi_A, \phi_B)}{k_\mathrm{B}T}=
	\sum_i \frac{\phi_i}{v_i}\ln(\phi_i) + 
	\sum_i e_i \phi_i + 
	\sum_{i,j}\frac{e_{ij}}{2}\phi_i\phi_j
	\;,
	\label{Eq:FreeEnergy}
	\end{align}
	where $k_\mathrm{B}$ is Boltzmann's constant and $i,j \in\{A,B,C\}$ using $\phi_C = 1 - \phi_A - \phi_B$.
	Here, the first term is the mixing entropy and the remaining terms capture enthalpic contributions~\cite{Flory_JCP_1944,Huggins_JCP_1941}.
	In particular, $e_i$ can be interpreted as internal energies, while $e_{ij}=e_{ji}$ capture interactions.
	Since $A$ molecules are soluble in the solvent, we for simplicity assume that they interact identically to the solvent ($e_{AA}=e_{AC}=e_{CC}$ and $e_{AB}=e_{BC}$).
	In the special case of a homogeneous system, the chemical potentials then read
	\begin{subequations}
		\label{Eq:chemPot}
		\begin{align}
		\mu_A &=k_\mathrm{B} T\left(
		\mo{A} + \ln\phi_A-\frac{v_A}{v_C}\ln\phi_C
		\right)
		\\
		\mu_B &= k_\mathrm{B} T\left(
		\mo{B} + \ln\phi_B-\frac{v_B}{v_C}\ln\phi_C
		- 2\chi \phi_B
		\right)
		\end{align}
	\end{subequations}
	where $2\chi=v_B(2e_{BC}-e_{BB}-e_{CC})$ is the Flory parameter capturing relevant interactions and $\mo{i}=1 - \frac{v_i}{v_C} + v_i(e_i - e_C + e_{iC} - e_{CC})$ quantifies internal energies for $i=A,B$.
	In non-homogeneous systems, $\mu_B$ would also contain a gradient term $\kappa\nabla^2 \phi_B$, where $\kappa = \kBT \, l^2\chi$ and $l$ determines the interface width~\cite{Mao_SoftMatter_2019}.
	\eqref{Eq:Contin} together with \eqref{Eq:chemPot} form a typical model for describing phase separation without chemical reactions.
	We first shortly discuss this classical case and then proceed to examine how different models for the reaction fluxes~$s$ affect the droplet formation.
	
	\subsection*{Amount of segregating material determines droplet size}
	
	\begin{figure}
		\includegraphics[width=\linewidth]{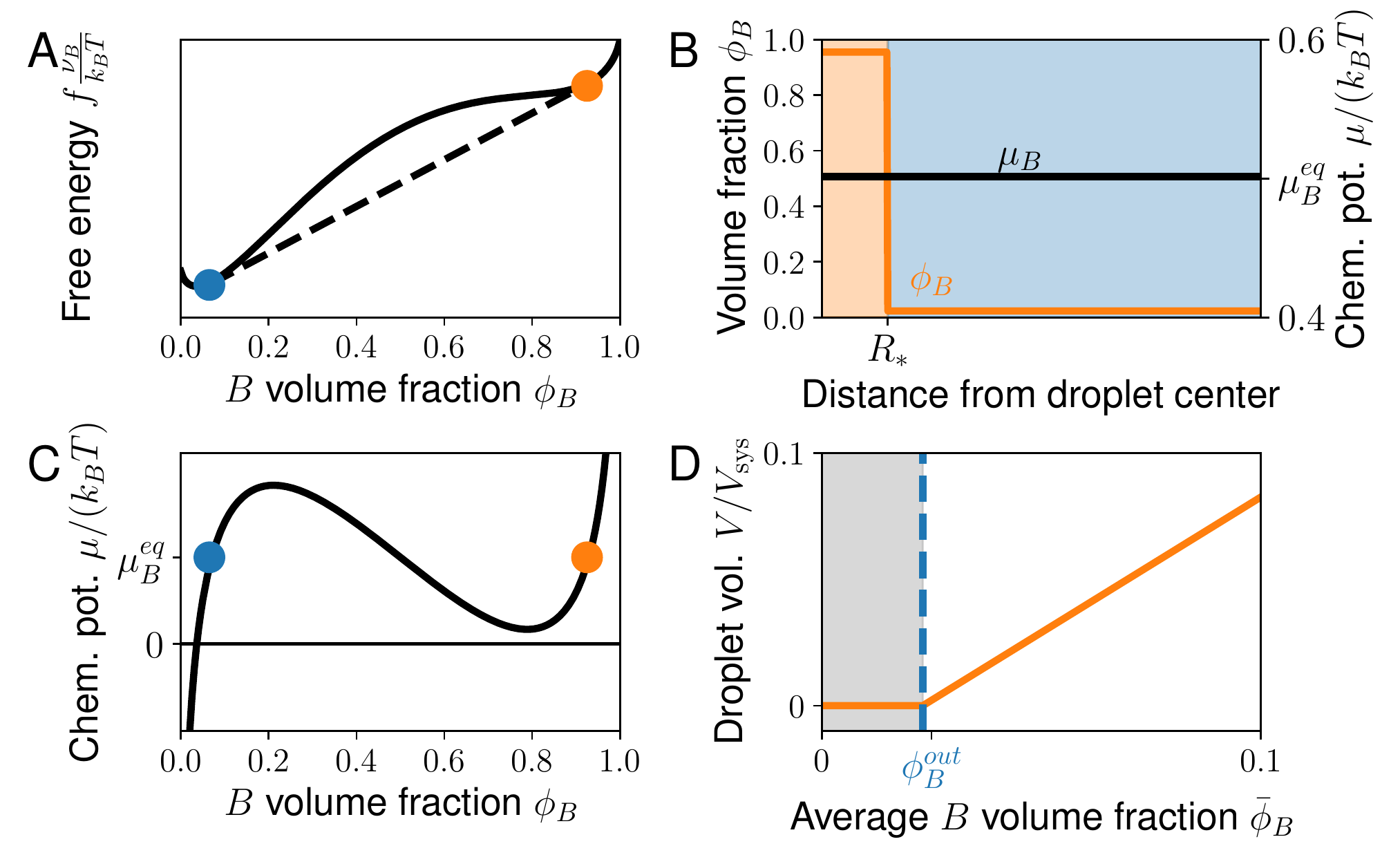}
		\caption{\textbf{The amount of droplet material controls the size of passive droplets.}
			\textbf{(A)}~Free energy density $f$ (Eq.\ref{Eq:FreeEnergy}) as a function of the volume fraction~$\phi_B$ (solid black line). A Maxwell construction (dashed black line) determines the fraction~$\phi_B^\ins$ inside droplets (orange circle) and the fraction~$\phi_B^\outs$ in the solvent (blue circle). 
			\textbf{(B)}~Equilibrium volume fraction~$\phi_B$ (left axis) and associated chemical potential~$\mu_B$ (right axis) as a function of the distance from the droplet center for a single droplet of radius $R_*$.
			\textbf{(C)}~Chemical potential~$\mu_B$ as a function of the volume fraction~$\phi_B$ of the droplet material (Eq.\ref{Eq:chemPot}).
			The chemical potentials are equal at the coexistence point (colored dots).
			\textbf{(D)}~Total volume~$\Vd$ of the droplet phase as a function of average fraction~$\bar{\phi}_B$ of the droplet material (Eq.\ref{Eq:VNR}).
			Droplets do not form for $\bar\phi_B < \phi_B^\outs$ (gray area).
			Model parameters are $\chi=3$, $w_B=0.5$, and $\bar{\phi}_B=0.026$ (in B).
		}
		\label{fig:A1}
	\end{figure}
	
	Without chemical reactions ($s=0$; see \figref{fig:A0}A), droplets can form when the free energy~$F$ of the demixed system is lower than that of the homogeneous system.
	This is the case if $\chi$ is large enough (see \figref{fig:A1}A) while the internal energies~$w_i$ are irrelevant since the total amount of each species is conserved~\cite{Weber_RepProgPhys_2019}.
	In equilibrium, the diffusive flux~$\bm{j}_B$ vanishes and the chemical potential~$\mu_B$ is homogeneous, while $\phi_B$ can vary strongly.
	The respective equilibrium fractions~$\phi_B^\ins$ and $\phi_B^\outs$ inside and outside of the droplet are given by a tangent construction (see \figref{fig:A1}A)~\cite{Mao_SoftMatter_2019}.
	They are constant and do not depend on the total composition of the system if the fraction of $A$ is small ($\phi_A \ll 1$).
	Without reactions, there are only two equilibrium states: Either everything is mixed or a single droplet enriched in $B$ forms.
	Even if multiple droplets form initially, e.g., due to nucleation, surface tension effects drive coarsening by Ostwald ripening~\cite{Ostwald_1897} or coalescence, so that all droplets merge into one~\cite{Weber_RepProgPhys_2019}.
	The volume~$\Vd$ of the droplet follows from material conservation and reads
	\begin{equation}
	\label{Eq:VNR}
	\Vd = \frac{\bar{\phi}_B-\phi_B^\outs}{\phi_B^\ins-\phi_B^\outs} \Vs 
	\;,
	\end{equation}
	where $\bar{\phi}_B = \Vs^{-1}\int \phi_B \, \mathrm{d}V$ is the average fraction of $B$ in the system.
	Note that the droplet can only exist when $\bar\phi_B > \phi_B^\outs$.
	All excess material beyond $\phi_B^\outs$ concentrates in the droplet, so $\Vd$ grows linearly with $\bar\phi_B$; see \figref{fig:A1}D.
	A biological cell can thus regulate whether a droplet exists and how large it gets by controlling the total amount of $B$.
	Protein amounts can be changed by production and degradation, although this is a costly and slow process.
	Moreover, $\Vd$ depends on the interaction parameter~$\chi$, which is a function of e.g. temperature, pressure, pH, and solvent composition.
	These parameters are either external to the cell or affect many other processes, so they are not ideal to regulate a specific droplet.
	Taken together, this analysis shows that additional processes are necessary to control droplets effectively.
	
	\subsection*{Chemical reactions control amount of segregating material}
	
	\begin{figure}
		\includegraphics[width=\linewidth]{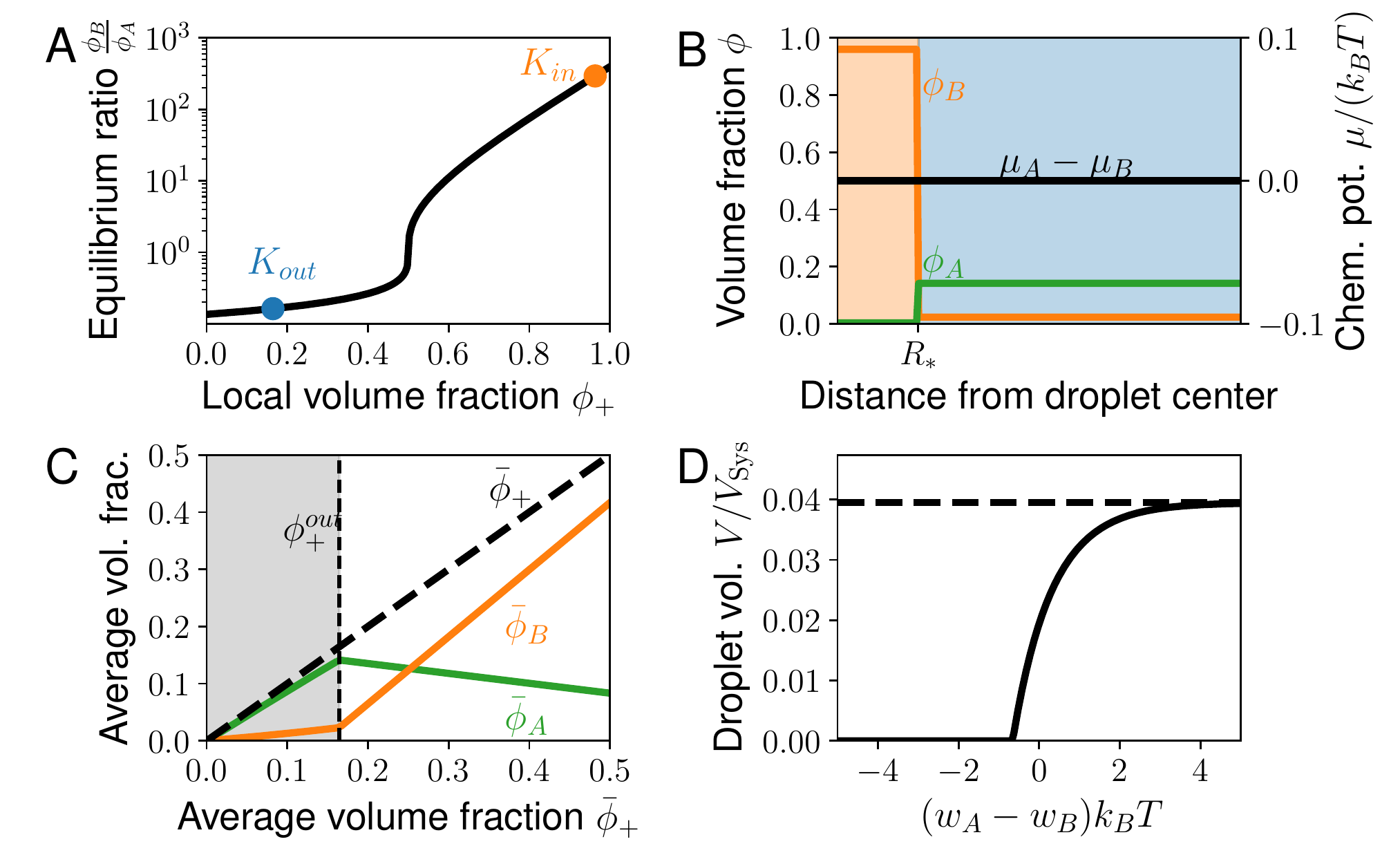}
		\caption{
			\textbf{Chemical equilibrium sets the amount of droplet material.}
			\textbf{(A)} The equilibrium ratio of the fractions of soluble and phase separating forms as a function of the total protein fraction $\phi_+=\phi_A+\phi_B$.
			\textbf{(B)} Volume fractions~$\phi_A$ and $\phi_B$ (left axis) and the associated chemical potential difference $\mu_A-\mu_B$ (right axis) as a function of the distance from the center of a droplet of radius~$R_*$.
			The composition of the droplet (orange shaded area) can differ strongly from the solvent (blue shaded area) even in equilibrium ($\mu_A=\mu_B$).
			\textbf{(C)} Average volume fractions of $A$ (green line) and $B$ (orange line) as a function of the total average protein volume fraction~$\bar\phi_+$ in the system.
			Without droplets (gray area), almost all protein is in the soluble~$A$ form, while the opposite is true for large droplets.
			\textbf{(D)} Total fraction $\Vd/\Vs$ occupied by droplets as a function of the difference $w_A-w_B$ between the internal energy of $A$ and $B$. 
			The dashed line marks the maximal volume, where all proteins are in form~$B$.
			Model parameters are $\chi=4$, $w_A-w_B=2$ (in A, B and C), $\bar{\phi}_+=0.2$ (in B) and $\bar{\phi}_+=0.06$ (in D)
		}
		\label{fig:A2}
	\end{figure}
	
	A chemical transition that modifies the physical properties of the droplet material can affect droplet formation.
	Our model captures this when we allow transitions between the soluble form~$A$ and the segregating form~$B$ of the material.
	The associated reaction rate $s$ is given by the difference of the rate $\sF$ of the forward reaction $A\rightarrow B$ and the rate $\sB$ of the opposite direction, $s=\sF-\sB$.
	In the simplest case, the transition $A\rightleftharpoons B$ does not require external energy input (see \figref{fig:A0}B), implying the detailed balance condition~\cite{Weber_RepProgPhys_2019}
	\begin{equation}\label{Eq:SpontReac}
	\frac{\sF}{\sB} = \exp\left(\frac{\mu_A-\mu_B}{k_\mathrm{B} T}\right)
	\;.
	\end{equation}
	Chemical equilibrium ($s=0$) is thus reached when $\mu_A = \mu_B$.
	
	In the simple case of a homogeneous system, the equilibrium state can be characterized by the fractions~$\phi_A^\mathrm{eq}$ and $\phi_B^\mathrm{eq}$ of the two forms.
	However, since the total fraction $\phi_+ = \phi_A + \phi_B$ of the component is conserved, it is convenient to also discuss the equilibrium constant $K=\phi_B^\mathrm{eq}/\phi_A^\mathrm{eq}$.
	Using the chemical equilibrium ($\mu_A=\mu_B$) and \Eqref{Eq:chemPot}, we find $K=\exp(w_A - w_B + 2\chi \phi_B)$, which shows that $K$ is strongly affected by the difference $w_A-w_B$ of the internal energies of $A$ and $B$.
	Note that $K$ is only a constant for an ideal solution ($\chi=0$).
	For non-ideal system, $K$ depends on the total fraction $\phi_+$, such that $K$ is larger when there is more material; see \figref{fig:A2}A.
	Taken together, this analysis shows that the chemical equilibrium depends on the environment.
	
	The only inhomogeneous equilibrium state of the system is again a single droplet enriched in $B$.
	The analysis of the homogeneous state implies that the ratio $\phi^\mathrm{eq}_B/\phi^\mathrm{eq}_A$ is larger inside the droplet than outside.
	This is because the droplet environment favors $B$ over $A$.
	Note that $A$ is enriched outside the droplet for the chemical potentials given by \Eqref{Eq:chemPot} (see \figref{fig:A2}B), but more general choices of $e_{ij}$ can enrich $A$ inside the droplet.
	For our system, $\phi_A$ is thus dilute in both phases in the common case that the system mostly consists of solvent ($\bar\phi_+ \ll 1$), where $\bar\phi_+=\Vs^{-1} \int \phi_+ \mathrm{d} V$ denotes the conserved total fraction of $A$ and $B$.
	In this case, we can determine the equilibrium fractions of $B$ from the tangent construction given in \figref{fig:A1}A since $\phi_A\ll1$ everywhere.
	We can then use $K_\ins=K(\phi_B^\ins)$ and $K_\outs=K(\phi_B^\outs)$ to determine the total fraction $\phi_+$ inside and outside the droplet, $\phi_+^\ins=(1+K_\ins^{-1})\phi_B^\ins$ and $\phi_+^\outs=(1+K_\outs^{-1})\phi_B^\outs$.
	The conservation of $\bar\phi_+$ then implies that the droplet volume~$\Vd$ is given by
	\begin{equation}\label{Eq:VRea}
	\Vd = \frac{\bar{\phi}_+-\phi_+^\outs}{\phi_+^\ins-\phi_+^\outs}\Vs
	\;.
	\end{equation}
	Similarly to the case without chemical reactions, a droplet can only form when $\bar\phi_+ > \phi_+^\outs$ and the total amount exceeding the threshold determines the droplet volume.
	However, the internal energy difference $w_A-w_B$ now also affect the droplet volume; see \figref{fig:A2}D.
	This is mainly because it changes the equilibrium constant $K_\outs$ and thus  $\phi^\outs_+$.
	Consequently, external parameters, like temperature, pH, etc., can now affect droplet formation also via the internal energies, thus allowing for a potentially stronger response.
	
	The chemical reactions clearly influence the droplet formation and thus the overall composition in the system.
	In particular, the relative amounts of $A$ and $B$ strongly depend on whether droplets form or not.
	\figref{fig:A2}C shows that the amount of $B$ in the system increases significantly when the total fraction~$\bar\phi_+$ exceeds the threshold $\phi_+^\outs$ so droplets form.
	
	We showed that the chemical transition allows for more detailed control of droplet formation.
	In particular, the droplet size depends on the total amount of protein, which can again be changed by the slow and costly processes of production and degradation.
	Additionally, the chemical transition affects droplet size since it determines the ratio of $A$ and $B$, which is mainly governed by the  energy difference $w_A - w_B$.
	These enthalpic parameters depend on global state variables, like temperature, pH, and salt, as well as on the protein sequence, which can be adjusted only on evolutionary time scales.
	Consequently, additional processes are necessary to have fast and specific control over droplets.

	\subsection*{Driven reactions allow enzymatic control of droplets}
	
	\begin{figure}
		\includegraphics[width=\linewidth]{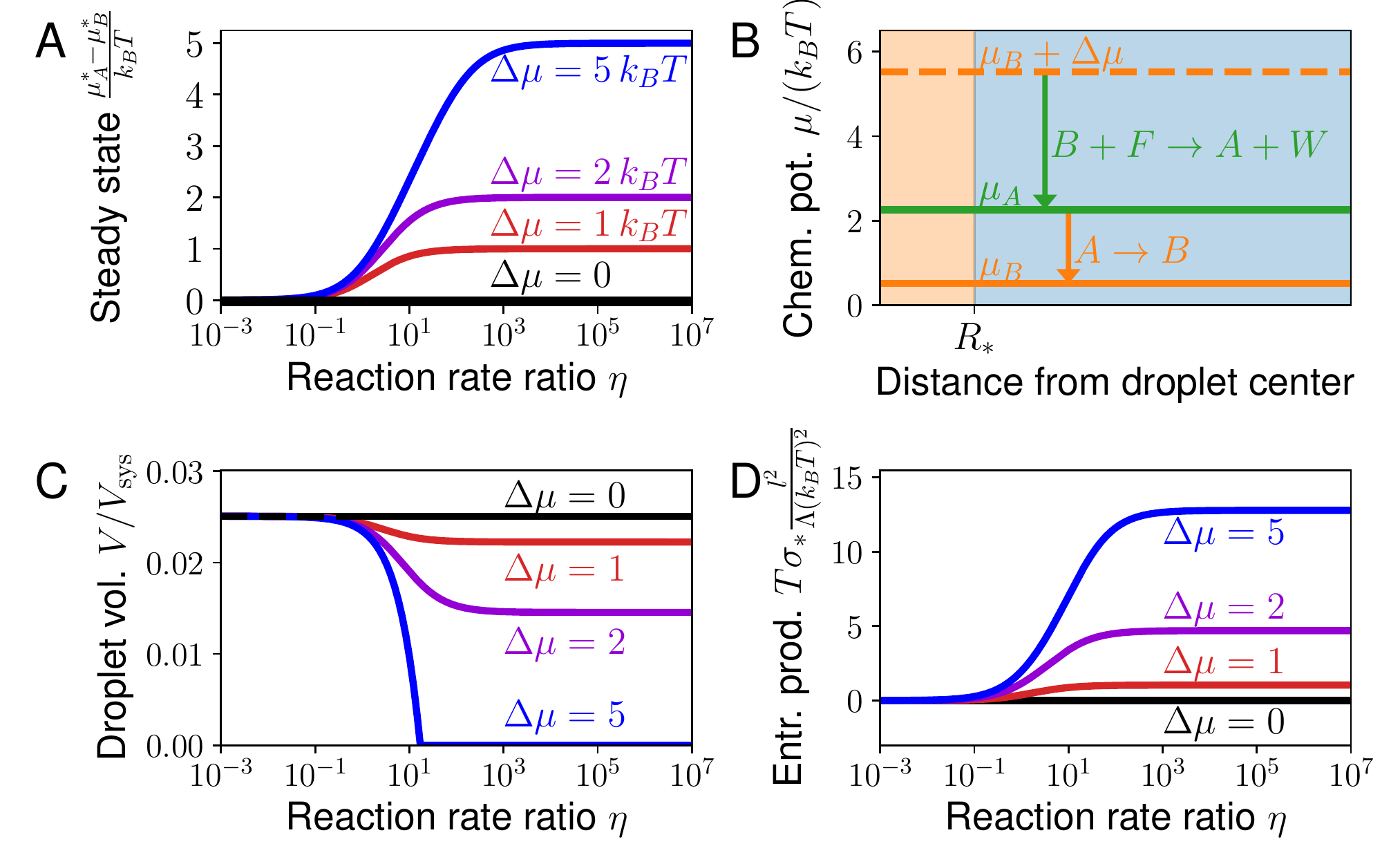}
		\caption{
			\textbf{A driven transition affects the chemical equilibrium.}
			\textbf{(A)} Chemical potential difference $\mu^*_A - \mu^*_B$ between $A$ and $B$ as a function of the ratio $\eta=\alpha_1^{-1}\alpha_2\exp(\mu_F/\kBT)$ of the reaction rate of passive and driven reaction in stationary state for several driving strengths~$\Delta\mu$.
			\textbf{(B)} Chemical potentials of $A$ (green line) and $B$ (solid orange line) as a function of the radial distance from the center of a droplet of radius~$R_*$.
			The dashed orange line marks the chemical potential difference of the driven reaction, indicating that it spontaneously produces~$A$ (green arrow), which is then turned back into $B$ by the passive reaction (orange arrow).
			Although the system is driven out of equilibrium, the reactions balance locally, so $\nabla\mu_i=0$.
			\textbf{(C)}
			Total fraction $\Vd/\Vs$ occupied by droplets as a function of~$\eta$ for different driving strengths~$\Delta\mu$.
			Droplets only vanish completely if the driving is strong enough.
			\textbf{(D)}~Entropy production rate~$T\sigma_*=-\Vs s_2\Delta\mu$ in the stationary state as a function of~$\eta$ for different driving strengths~$\Delta\mu$. 
			Model parameters are $\chi=3.5$, $w_A-w_B=3$, and $\phi_+=0.06$
		}
		\label{fig:A3}
	\end{figure}
	We next extend our system by allowing the transition $A\rightleftharpoons B$ to also be driven by an external energy input; see \figref{fig:A0}C.
	In particular, we introduce a second reaction, $A + W \rightleftharpoons B + F$, where $F$ and $W$ respectively denote fuel and waste molecules, e.g. ATP and ADP.
	For simplicity, we consider the case where $F$ and $W$ are dilute and homogeneously distributed, so they do not affect phase separation directly.
	To keep the system away from equilibrium, we assume that the chemical potential difference $\Delta\mu = \mu_F-\mu_W>0$ is constant, e.g., because of ATP regeneration.
	Taken together, $\Delta\mu$ can be interpreted as an ubiquitous external energy source.
	
	The driven reaction obeys the detailed balance condition~\cite{Weber_RepProgPhys_2019}
	\begin{equation}
	\frac{\sF_2}{\sB_2} = 
	\exp\left(\frac{\mu_A-\mu_B-\Delta\mu}{k_\mathrm{B} T}\right)
	\;,\label{Eq:ReactDriv}
	\end{equation}
	where $\sF_2$ and $\sB_2$ are the respective forward and backward rates.
	The net rate $s_2=\sF_2 - \sB_2$ of the driven reaction thus vanishes when $\mu_A-\mu_B=\Delta\mu$.
	This condition is incompatible with the chemical equilibrium of the passive reaction, $\mu_A=\mu_B$, discussed in the previous section.
	This implies that the driven system cannot reach thermodynamic equilibrium.
	
	To understand the behavior of the driven system, we first consider stationary states where the total reactive flux, $s=s_1+s_2$, vanishes.
	Here, $s_1 = \sF_1 - \sB_1$ is the rate associated with the passive reaction, where $\sF_1$ and $\sB_1$ obey the detailed balanced condition given by \Eqref{Eq:SpontReac}.
	Taken together, the condition $s=0$ requires that the stationary state chemical potentials $\mu_A^*$ and $\mu_B^*$ obey
	\begin{align}
	\mu_A^* - \mu_B^* = \Delta\mu -  k_BT\ln\left[
	\frac{\exp(\Delta\mu/k_\mathrm{B}T) + \eta}{1 + \eta}
	\right] \;,
	\label{Eq:NESS}
	\end{align}
	where $\eta = \sB_2/\sB_1$ is the ratio of the backward reaction rates.
	Note that this condition corresponds to the passive case ($\mu_A^* = \mu_B^*$) for $\eta=0$, while the driven reaction dominates ($\mu_A^* - \mu_B^* = \Delta\mu$) for $\eta\rightarrow\infty$; see \figref{fig:A3}A.
	In general, we have $0 \le \mu_A^* - \mu_B^* \le \Delta\mu$, so that the passive reaction creates the segregating form $B$ while the driven reaction destroys it.
	
	In general, the backward rates~$\sB_i$ depend on composition, since they describe the kinetics of the chemical reactions~\cite{Carati_1997}.
	A simple model for chemical reactions is transition state theory~\cite{hanggi_revmodphys_1990,Pagonabarraga_PhysA_1997}, where the forward and backward rates only depend on the chemical potentials of the reactants and products, respectively.
	Using this theory, we find $\sB_1 = \alpha_1 \exp(\mu_B/\kBT)$ and $\sB_2=\alpha_2\exp[(\mu_B + \mu_F)/\kBT]$, where $\alpha_i$ are constant pre-factors that can be influenced by enzymes; see SI.
	This implies that $\eta=\alpha_2 \alpha_1^{-1} \exp(\mu_F/\kBT)$, and thus also $\mu_A^* - \mu_B^*$, are constant (see \Eqref{Eq:NESS}).
	Taken together with \Eqref{Eq:Contin}, we thus find that all stationary states with $s=0$ must have homogeneous chemical potentials; see \figref{fig:A3}B.
	
	The driven system can be mapped to the system with passive reactions by altering the internal energies, $w_B \mapsto w_B + (\mu_A^* - \mu_B^*)/k_\mathrm{B} T$.
	Consequently, this system possesses the same stationary states as the passive system, so that at most a single droplet can form and its volume is given by \Eqref{Eq:VRea}.
	However, the driven chemical reaction can now be used to control the droplet volume, e.g., by enzymatic activity.
	For instance, an increased activity of an enzyme that catalyzes the driven reaction corresponds to an increase in $\alpha_2$.
	This results in an increase of $\eta$, $\mu_A^*-\mu_B^*$, $s$, $K^{-1}$, and $\phi_+^\outs$, which leads to a smaller droplet volume $V$; see \figref{fig:A3}C.
	Equivalently, raising the external potential~$\Delta\mu$ also reduces~$V$.
	In particular, any change that increases $\phi_+^\outs$ beyond the average fraction $\bar\phi_+$ of available material will dissolve all droplets.
	Note that this dissolution by enzymatic reactions happens without degrading the material, so droplets could re-form quickly when the original conditions are restored.
	However, the potential for this quick response comes at the energetic cost, quantified by the entropy production rate (see \figref{fig:A3}D), of keeping the droplets dissolved~\cite{Wurtz_PRL_2018}.
	
	We showed that at most a single droplet can be stable when the net flux of the chemical reactions vanish everywhere ($s=0$).
	To also regulate the droplet count, we thus need inhomogeneous states where $s\neq0$.
	However, we show in the SI that there are no stationary states with $s\neq 0$ if $\eta$ and $\Delta\mu$ are the same everywhere.
	Consequently, $\eta$ or $\Delta\mu$ must vary in space to have multiple stable droplets.
	This could be achieved by imposing spatial heterogeneity, e.g., by producing the fuel or enriching enzymes at particular locations, which would be reflected in the droplet arrangement.
	Alternatively, the fuel or enzymes can segregate into the droplets spontaneously, which is observed experimentally~\cite{Boehning_nature_2018}.
	
	\subsection*{Segregated enzymes can control droplet size and count}
	The main idea to control droplets is to use an enzyme that regulates the chemical transition and segregates into droplets.
	As an example, we consider an enzyme $E$ that affects the driven reaction; see \figref{fig:A0}D.
	In the simplest case, the rate $s_2$ of this reaction is proportional to the volume fraction~$\phi_E$ of the enzyme,
	\begin{equation}\label{Eq:EnzymaticReaction}
	s_2 = \alpha^E_{2} \phi_E
	\left[\exp\left(\frac{\mu_A +\mu_W}{\kBT}\right)- \exp\left(\frac{\mu_B + \mu_F }{\kBT}\right)\right]
	\;,
	\end{equation}
	which follows from \Eqref{Eq:ReactDriv} and transition state theory; see SI.
	Here, $\alpha^E_{2}$ is a constant pre-factor, so that this case is equivalent to the one discussed in the previous section if $\phi_E$ is homogeneous.

	The distribution of the enzyme will be inhomogeneous if it segregates into droplets.
	We model this by introducing an additional Flory parameter~$\chi_E$, which describes the interaction of the enzyme with the other components; see SI.
	For simplicity, we consider dilute enzyme concentrations, so it does not  affect the phase separation significantly.
	Consequently, $\chi_E$ controls how strongly the enzyme segregates into the droplet~\cite{Weber_elife_2019},
	\begin{equation}
	\frac{\phi_E^\ins}{\phi_E^\outs}
	\approx e^{\chi_E(\phi_B^\ins-\phi_B^\outs)}
	\;;
	\end{equation}
	see SI.
	In particular, the enzyme is homogeneously distributed for $\chi_E=0$, corresponding to the case discussed in the previous section. 
	
\begin{figure}
		\includegraphics[width=\linewidth]{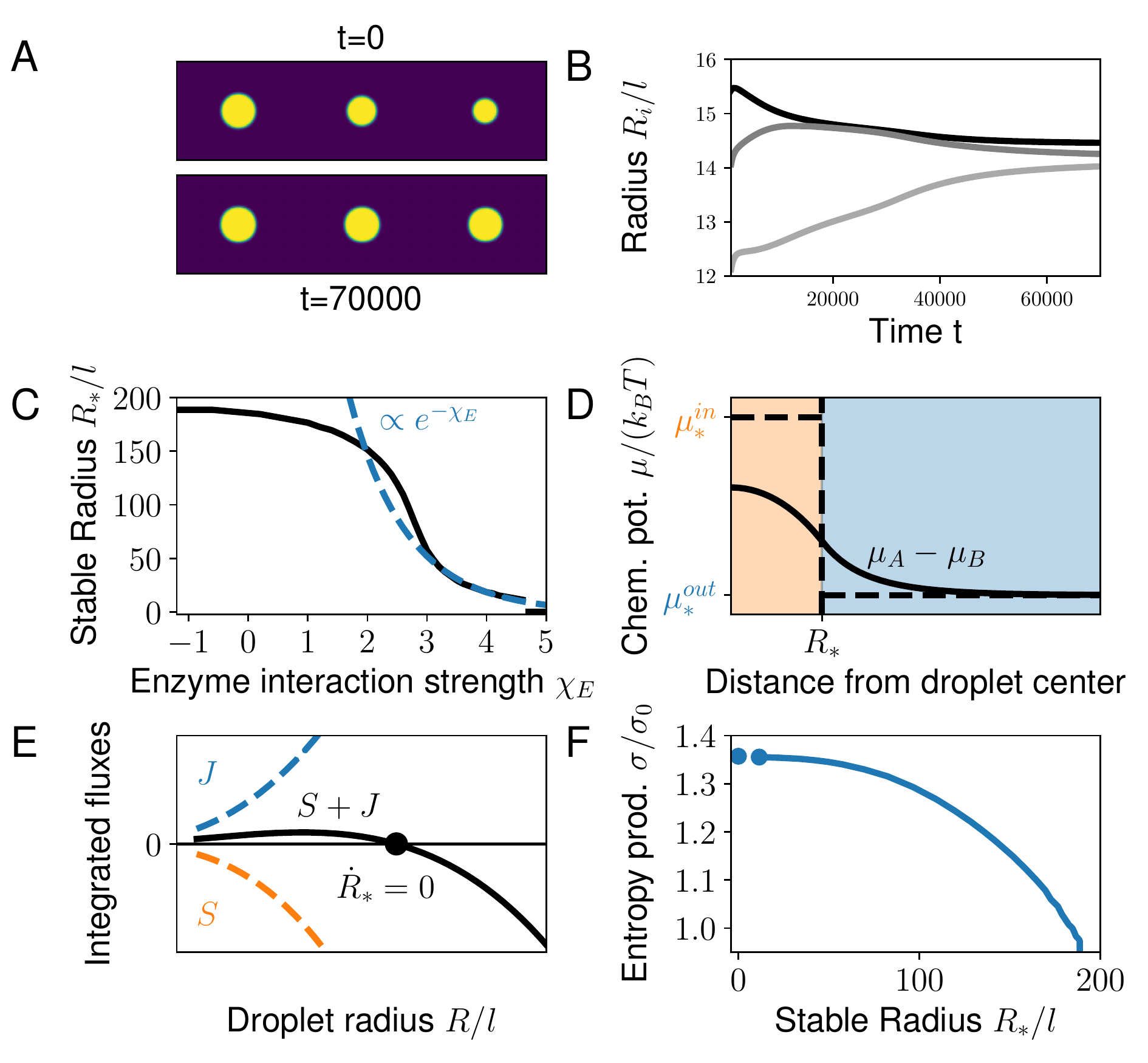}
		\caption{\textbf{A segregated enzyme can control and stabilize multiple droplets.}
			\textbf{(A)}~Snapshots of a numerical simulation of three droplets in a cylindrical geometry at two time points.
			Simulation parameters are $\chi=\chi_E=4$, $w_A-w_B=2$, $\Delta\mu=10\kBT$, $\Lambda=1$, $\alpha_1=10^{-3}\,\Lambda/l^2$, $\eta=3$, and $\bar{\phi}_+=0.25$.
			\textbf{(B)} Droplet radii~$R_i$ of the simulation in (A) as a function of time.
			\textbf{(C)} Stable droplet radius $R_*$ as a function of the interaction parameter~$\chi_E$ for the enzyme.
			\textbf{(D)} The steady state chemical potential difference between $A$ and $B$ as a function of the distance from the droplet center.
			The chemical potential gradient does not vanish across the interface between droplet (orange) and solvent phase (blue), implying diffusive fluxes.
			Far away from the droplet, the reactions cancel each other ($s=0$), while $s<0$ in the droplet and $s>0$ in the solvent close to the interface.    
			\textbf{(E)} The reaction flux in the droplet (orange dashed line) and the solvent (blue dashed line) as well as their sum (black dashed line) as a function of the droplet radius.
			The two fluxes are equal and opposite at the stationary state (black dot).
			\textbf{(F)}~Entropy production rate~$T\sigma=-\int s_2\Delta\mu \,\mathrm{d}V$ associated with (C) as a function of $R_*$ normalized to the entropy production $\sigma_0$ for $\chi_E=0$.
			Model parameters used in (C--F) are $\chi=4$, $\bar{\phi}_+=0.06$, $\bar{\phi}_E=0.001$, $\Lambda=1$, $\alpha_1=5\cdot10^{-4} \,\Lambda/l^2$, $\eta = 1/5$, $w_A-w_B=4$, and $\Delta\mu=5\kBT$.
		}
		\label{fig:A4}
	\end{figure}

	The enzyme is enriched in the droplet when $\chi_E>0$.
	In this case, the driven reaction can stabilize multiple droplets at the same size; see \figref{fig:A4}A--B.
	To understand this behavior, we analyze a single droplet in a large system.
	\figref{fig:A4}D shows that the chemical potentials of $A$ and $B$ are now inhomogeneous even in the stationary state.
	This implies diffusive fluxes, which are driven by the non-equilibrium chemical reactions:
	Effectively, inside the droplet, the driven chemical reaction turns the segregating form~$B$ into the soluble form~$A$, while form~$A$ transitions back to $B$ spontaneously outside.
	The resulting imbalances between the inside and the outside are compensated by the diffusive fluxes.
	Consequently, the chemical reactions drive a cycle of diffusive fluxes; see \figref{fig:A0}D.
	
	The numerical simulations also show that the stable droplet radius~$R_*$ decreases with larger enzyme segregation (larger $\chi_E$); see \figref{fig:A4}C.
	In the stationary state, the diffusive influx~$J$ of $B$ toward the droplet is balanced by the reactive flux~$S$ of $B \rightarrow A$ inside the droplet ($J=S$).
	In the simplest case, $J$ is diffusion limited, $J\approx a_1 R$, while the reaction is homogeneous in the droplet, implying $S\approx a_3 \exp(\chi_E) R^3$; see SI.
	Consequently, the stable radius scales as $R_* \sim \exp(-\frac12\chi_E)$.
	In more realistic cases, the reaction affects the influx~$J$, leading to $J\approx a_2 R^2$, which implies $R_* \sim \exp(-\chi_E)$; see \figref{fig:A4}E.
	In both cases, a stable droplet size exists when the fluxes $J$ and $S$ are equal; see \figref{fig:A4}E.
	
	The droplet size regulation depends on the non-equilibrium chemical reactions, which maintain a chemical potential difference between the droplet and its surrounding; see \figref{fig:A4}D.
	We observe that the associated entropy production rate~$\sigma$ increases for smaller radii~$R_*$; see \figref{fig:A4}F.
	This suggests that keeping droplets small consumes more fuel~$F$.
	In particular, preventing droplet formation ($R_*=0$) is costly.
	Conversely, larger droplets require smaller entropy production, although it is still non-zero, similar to the case in the previous section.
	The fact that droplets reach a stable size implies that multiple droplets can coexist in a larger system.
	Since all droplets attain the same volume~$V_*$, the number~$N$ of droplets is simply $N=V/V_*$, where the total volume $V$ of the droplet phase can be approximated by \Eqref{Eq:VRea}.
	In particular, $V$ depends on the total fraction~$\bar\phi_+$ of $A$ and $B$, the Flory parameter~$\chi$, the internal energies $w_A-w_B$, the driving strength $\Delta\mu$, and the reaction rate ratio~$\eta$.
	Conversely, the stable radius $V_*$ is additionally controlled by $\chi_E$, so the droplet count~$N$ and the individual volume~$V_*$ can be adjusted independently.

	\section*{Discussion}
	
	We introduced a model that explains how chemical reactions can control liquid-like droplets.
	In particular, we identified three ingredients necessary for effective size control: (i) The chemical modification of the droplet material must convert it to a soluble form, (ii) this modification must involve a driven reaction using a chemical fuel, and (iii) the reaction dynamics must differ inside and outside the droplet, e.g., by localizing enzymes appropriately.
	The fuel, combined with the imbalance of the reaction, maintains a chemical potential difference between the inside and the outside, which results in sustained diffusive fluxes.
	This effectively removes droplet material from the droplet, while producing it outside, which explains the stable size of this externally-maintained droplet~\cite{Weber_RepProgPhys_2019}.
	In an alternative interpretation, the enzymes enriching in the droplet inhibit further growth, which we already identified as a common motif for size control in biological cells~\cite{Soeding_TCB_2020}.

	The stable droplet size~$R_*$ predicted by our model is mainly governed by the chemical transition inside the droplet; see SI.
	In particular, $R_* \sim [3 D \Delta c/(k_0 c_E^\mathrm{in})]^{1/2}$ where
	$D\approx \unitfrac[1]{\upmu m^2}{s}$ is a typical diffusivity~\cite{Schavemaker_2018}, while the other parameters can vary widely~\cite{Milo_CellBio_2015}.
	Here, $\Delta c$ quantifies the concentration variation of droplet material $B$ in the dilute phase, $k_0$ is the catalytic rate constant, and $c_E^\mathrm{in}$ is the enzyme concentration inside the droplet.
	For strong reactions ($k_0 \approx \unit[100]{s^{-1}}$) and strong enzyme segregation ($\Delta c/c_E^\mathrm{in} \approx 0.1$), we find very small droplets ($R_* \approx \unit[0.05]{\upmu m}$).
	Conversely, droplets are much larger ($R_* \approx \unit[17.3]{\upmu m}$) for weaker reactions ($k_0 \approx \unit[0.1]{s^{-1}}$) and moderate segregation ($\Delta c/c_E^\mathrm{in} = 10$).
	Consequently, droplets can be stabilized on all length scales relevant to biological cells.
	In particular, $R_*$ is governed by intrinsic model parameters and is thus independent of system size, similar to other theoretical predictions from combining phase separation with chemical reactions~\cite{Carati_1997,Wurtz_PRL_2018,Zwicker_PRE_2015,li_non-equilibrium_2020}.
	In all these systems, Ostwald ripening is suppressed and multiple droplets can stably coexist.
	The droplet count is given by the ratio of the total amount of droplet material and the stable droplet volume.

	Whether droplets form and how large they get mainly depends on the available amount of droplet material.
	Our model reveals that this key quantity can be regulated on many time scales in biological cells:
	Adapting the genetic sequence on evolutionary time scales affects the internal energies of the soluble and segregating forms, thus influencing the fraction of droplet material; see \figref{fig:A2}D.
	On the time scale of minutes to hours, protein production and degradation affects the overall composition; see \figref{fig:A1}D.
	Faster time scales are accessible using active processes:
	By activating and deactivating enzymes, the cell can regulate the reaction rates~$\alpha_1$ and $\alpha_2$ and thus the balance between the two forms.
	Note that even small changes in these rates have a huge impact on droplet size when the driving~$\Delta\mu$ is sufficiently strong; see \figref{fig:A3}C.
	This active regulation allows cells to quickly adapt their biomolecular condensate in response to internal and external signals~\cite{alberti_JMolBio_2018,zhao_DevCell_2020,jaqaman_CurrOpCellBio_2021}.
	However, this flexibility comes at the cost of continuous turnover of droplet material, reminiscent of enzymatic futile cycles~\cite{Samoilov2005}.
	Moreover, the continuous turnover could prevent the observed aging of biomolecular condensates~\cite{alberti_annrevgen_2019,jawerth_protein_2020,Linsenmeier2021}.
	
	Our model unveils the required ingredients for droplet size regulation since it obeys thermodynamic constraints, in contrast to our earlier theory~\cite{Zwicker_PRE_2015}.
	Similar to electro-chemical systems~\cite{bazant_faraday_2017}, the chemical reactions in our system cannot be described by the law of mass action since phase separating solutions are non-ideal.
	In particular, the associated equilibrium constants differ inside and outside the droplet; see \figref{fig:A2}A.
	To investigate this further, our theory could be extended to client chemical reactions~\cite{Weber_elife_2019}, multi-component droplets~\cite{riback_nature_2020}, complex multi-layered droplets~\cite{Mao_SoftMatter_2019,mao_designing_2020,swain_biosoctrans_2020}, and multiple different droplets affecting each other~\cite{hondele_nat_2019}, which are all relevant in biological cells.
	Moreover, it is unclear how active droplets interact with other sub-cellular structures, like the cytoskeleton~\cite{wiegand_emtoplife_2020}, or generally with the elastic properties of their surrounding~\cite{rosowski_elastic_2020,vidal-henriquez_theory_2020}.
	It will be interesting to test our ideas with engineered condensates~\cite{bracha_natbiotech_2019} using fueled chemical reactions~\cite{donau_natcom_2020}.

	\section*{acknowledgment}
	We thank Johannes Söding, Estefania Vidal, Christoph A. Weber, and Noah Ziethen for helpful discussions and Lucas Menou for critically reading the manuscript.
	This work has been supported by the Max Planck Society.

	
	\bibliographystyle{unsrt}

	\setcounter{equation}{0}
	\setcounter{section}{0}
	\renewcommand{\theequation}{S.\arabic{equation}}
	\newpage
	\onecolumngrid
	
	\section*{Supplementary Information}
	
	\section{Multicomponent free energy}
	As a basis for our discussion we use a multicomponent free energy $F$, which depends on the volume fractions $\Phi=(\phi_1,...,\phi_N)$ of all $N$ components according to
	\begin{align}\label{Sieq:F}
	F[\Phi] = \int \biggl[
	f(\Phi)
	-\!\!\!\!\sum_{i,j=1}^N\frac{\kappa_{ij}}{2}\nabla\phi_i . \nabla\phi_j
	\biggr] \mathrm{d}V \;.
	\end{align}
	Here $f(\phi)$ is the free energy density and the $\kappa_{ij}=\kappa_{ji}$ terms includes longer range interaction between species $i$ and $j$, penalizing concentration gradients\cite{Weber_RepProgPhys_2019}.
	We use a Flory-Huggins like free energy density given by
	\begin{align}\label{Sieq:FH}
	\frac{f(\Phi)}{k_\mathrm{B}T}=
	\sum_i \frac{\phi_i}{v_i}\ln(\phi_i) + 
	\sum_i e_i \phi_i + 
	\sum_{i,j}\frac{e_{ij}}{2}\phi_i\phi_j
	\;,
	\end{align}
	where the first term is the entropic contribution with constant molecular volume $v_i$, $e_i$ is the internal energy and $e_{ij}=e_{ji}$ is the enthalpic interaction energy between species $i$ and $j$.\\
	We discuss incompressible systems only, so $\sum_i \phi_i=1$ at all times. Using this, we can rewrite the interaction term according to
	\begin{align}
	\sum_{i,j}\frac{e_{ij}}{2}\phi_i\phi_j=\sum_i \frac{e_{ii}}{2}\phi_i^2 + \sum_{j\neq i}\frac{e_{ij}}{2}\phi_i\phi_j=\sum_i \frac{e_{ii}}{2}\phi_i + \sum_{j\neq i}\frac{e_{ij}-e_{ii}}{2}\phi_i\phi_j
	\end{align}
	and using $2\chi_{ij} = 2e_{ij}-e_{ii}-e_{jj}$, we arrive at
	\begin{align}\label{Sieq:fed}
	\frac{f(\Phi)}{\kBT} = \sum_i\frac{\phi_i}{v_i}\ln(\phi_i) + \sum_i (e_i+\frac{e_{ii}}{2})\phi_i + \sum_{i,j}\frac{\chi_{ij}}{2}\phi_i\phi_j.
	\end{align}
	Here $\chi_{ij}$ shows if species $i$ and $j$ prefer to mix ($\chi_{ij}<0$) or demix ($\chi_{ij}>0$) due to enthalpic interactions (where $\chi_{ii}=0$ by definition). Furthermore, we assume that the gradient term in eq.\ref{Sieq:F} is determined by the same interactions as the local free energy, such that $\kappa_{ij}=\kBT l^2\chi_{ij}$, where $l$ is the characteristic length scale of the interaction\cite{Mao_SoftMatter_2019}.\\
	Using Eq.\ref{Sieq:F} and \ref{Sieq:fed}, we get the chemical potential according to $\bar{\mu}_i = \frac{\delta F}{\delta N_i}$, which results in
	\begin{align}\label{Sieq:CP}
	\frac{\bar{\mu}_i}{\kBT} = 1+\ln(\phi_i) + v_i(e_i+\frac{e_{ii}}{2}) + \sum_{j=1}^N v_i\chi_{ij}\phi_j.
	\end{align}
	The incompressibility condition $1=\sum_i \phi_i$ implies that the diffusion of particles is connected, which we use to replace the $N$-th species by $\phi_N = 1-\sum_{i\neq N}\phi_i$. Therefore diffusive fluxes are not driven by the chemical potentials given in Eq.\ref{Sieq:CP} but by exchange chemical potentials
	\begin{align}
	\mu_i = \bar{\mu}_i-\frac{v_i}{v_N}\bar{\mu}_N,
	\end{align}
	which describe the energy change when $i$ particles are replaced by the same volume of $N$ particles.\\
	For a ternary fluid with $i=A,B,C$ and $N=C$, this results in
	\begin{subequations}
		\begin{align}
		\frac{\mu_A}{\kBT} &= 1-\frac{v_A}{v_C} + v_A(e_A-e_C+e_{AC}-e_{CC}) + \ln(\phi_A)-\frac{v_A}{v_C}\ln(\phi_C) + \sum_{j\neq C}v_A(\chi_{Aj} - \chi_{AC}-\chi_{Cj})\phi_j,\\
		\frac{\mu_B}{\kBT} &= 1-\frac{v_B}{v_C} + v_B(e_B-e_C+e_{BC}-e_{CC}) + \ln(\phi_B)-\frac{v_B}{v_C}\ln(\phi_C) + \sum_{j\neq C}v_B(\chi_{Bj} - \chi_{BC}-\chi_{Cj})\phi_j.
		\end{align}
	\end{subequations}
	Now for the special case where $B$ phase separates from $A$ and $C$ and $A$ interacts similar to $C$, such that $\chi_{AC}=0$ and $\chi_{AB} =\chi_{BC} =\chi/v_B$, we end up with
	\begin{subequations}
		\label{Sieq:CP2}
		\begin{align}
		\frac{\mu_A}{\kBT} &= w_A + \ln(\phi_A)-\frac{v_A}{v_C}\ln(\phi_C),\\
		\frac{\mu_B}{\kBT} &= w_B + \ln(\phi_B)-\frac{v_B}{v_C}\ln(\phi_C) -2\chi\phi_B,
		\end{align}
	\end{subequations}
	with $w_A = 1-v_A/v_C + v_A(e_A-e_C)$ and $w_B = 1-v_B/v_C + v_B(e_B - e_C + e_{BC}-e_{CC})$, which is the form we use in the main text.
	
	\section{Enzyme segragation}
	Here we discuss how a dilute enzyme $E$ distributes in the two phases discussed in the main text. The discussion is similar to the derivation of the segragation coefficient in \cite{Weber_elife_2019}. Therefore we extend the free energy introduced above by a fourth component $\phi_E\ll 1$.\\
	Assuming $\phi_B\gg \phi_E$ the enzymes exchange chemical potential takes the form
	\begin{align}
	\frac{\mu_E}{\kBT} \approx w_E + \ln(\phi_E) - \frac{v_E}{v_C}\ln(\phi_C) - (\chi_E + (v_E/v_B)\chi)\phi_B, 
	\end{align}
	where $\chi_E = v_E(\chi_{EC}-\chi_{EB})$ determines whether it is energetically favorable for the enzyme to be in the $B$ rich droplet phase ($\chi_E>0$) or the dilute phase ($\chi_E<0$).\\
	The maxwell construction discussed in Fig.2 of the main text results in $\phi_B = \phi_B^\ins/\phi_B^\outs$ for droplet and solvent phase respectively, with chemical potentials $\mu_B(\phi_B^\ins)=\mu_B(\phi_B^\outs)=w_B-\chi$ for $\phi_A\ll 1$ and $v_A=v_B=v_C=v$. Those assumptions allow analytical calculations, but are not necessary and the results apply more generally. With this equal enzyme chemical potential in both phases leads to
	\begin{align}
	\ln\left(\frac{\phi_E^\ins}{\phi_E^\outs}\right) -\chi_E(\phi_B^\ins - \phi_B^\outs) = \frac{v_E}{v}\left(\ln\left(\frac{\phi_C^\ins}{\phi_C^\outs}\right) + \chi (\phi_B^\ins - \phi_B^\outs)\right).
	\end{align}
	For $v_C=v_B$ and $\phi_A\ll 1$, the free energy is symmetric, such that $\phi_B^\ins+\phi_B^\outs = 1$ and $\phi_C\approx 1-\phi_B$. In this case
	\begin{align}
	\ln\left(\frac{\phi_C^\ins}{\phi_C^\outs}\right) + \chi (\phi_B^\ins - \phi_B^\outs) \approx \ln\frac{\phi_B^\outs}{1-\phi_B^\outs} + \chi (1 - 2\phi_B^\outs) = \mu_B(\phi_B^\outs) - w_B + \chi =0,
	\end{align}
	such that
	\begin{align}
	\frac{\phi_E^\ins}{\phi_E^\outs} = \exp(\chi_E(\phi_B^\ins - \phi_B^\outs))=\Gamma.
	\end{align}
	So the segragation is mostly determined by the strength of phase separation $\phi_B^\ins - \phi_B^\outs$ and the different interaction of the enzyme with $B$ and $C$, $\chi_E$. In the strong phase separation regime, we approximate $\phi_B^\ins-\phi_B^\outs\approx 1$ and thus
	\begin{align}
	\frac{\phi_E^\ins}{\phi_E^\outs} = \exp(\chi_E),
	\end{align}
	which is the form we use in the main text for simplicity.
	So far we have calculated the relative amount of enzyme in droplet and solvent phase. No we calculate the actual volume fraction in each phase for which we need the average volume fraction, $\bar{\phi}_E$ and the total fraction of droplet phase $\Vd/\Vs$. Because the total amount of enzyme is conserved and the volume fraction in droplet and solvent phase are constant, we get
	\begin{align}
	\Vd \phi_E^\ins + (\Vs - \Vd)\phi_E^\outs = \Vs \bar{\phi}_E,
	\end{align}
	and thus 
	\begin{subequations}
		\begin{align}
		\phi_E^\outs &= \frac{1}{1-\Vd/\Vs(1-\Gamma)}\bar{\phi}_E\approx \left(1+(1-\Gamma)\frac{\Vd}{\Vs}\right)\bar{\phi}_E,\\
		\phi_E^\ins &= \frac{\Gamma}{1-\Vd/\Vs(1-\Gamma)}\bar{\phi}_E\approx \left(1+(1-\Gamma)\frac{\Vd}{\Vs}\right)\Gamma\bar{\phi}_E.
		\end{align}
	\end{subequations}
	For droplet size control we usually assume $\Gamma\gg1$ and $\Vd/\Vs\ll 1$, which can be used to write $\phi_E^\outs\approx \bar{\phi}_E$ and $\phi_E^\ins\approx \Gamma\bar{\phi}_E$ in a first order approximation for the analytical calculations.

	\section{Detailled balance of the rates}
	For a given reaction converting reactants $R_i$ into products $P_i$, $R_i\rightleftharpoons P_i$, detailed balance relates the forward $\sF$ and backward $\sB$ reaction rate according to
	\begin{equation}
	\frac{\sF}{\sB} = \exp\left[\frac{\sum_i \mu_{R_i}-\sum_i\mu_{P_i}}{\kBT}\right].
	\end{equation}
	Where $\mu_i$ is the chemical potential of species $i$. The total reaction flux $s=\sF-\sB$ then vanishes in equilibrium where $\sum_i \mu_{R_i}=\sum_i \mu_{P_i}$.
	But detailed balance does not tell us how forward and backward rate look like individually. Here we assume that the forward rate is a function of the forward chemical potential $\mu_F = \sum_i \mu_{R_i}$, $\sF(\mu_F)$, and the backward rate is a function of the backward chemical potential $\mu_B = \sum_i \mu_{P_i}$, $\sB(\mu_B)$.
	This approximation has the advantage that forward and backward rate have a symmetric form and they depend on the energetics of the species that take part in forward or backward reaction only.
	In this general form we can rewrite the detailed balance conditiom according to
	\begin{equation}
	\sF(\mu_F)\left(-\frac{\mu_F}{\kBT}\right) = \sB(\mu_B)\left(-\frac{\mu_B}{\kBT}\right).
	\end{equation}
	Assuming that $\mu_F$ and $\mu_B$ can in general be varied independently, both sides of the equation have to be constant individually and thus
	\begin{subequations}
		\begin{align}
		\sF(\mu_F) = \alpha\exp\left(\frac{\mu_F}{\kBT}\right),\\
		\sB(\mu_B) = \alpha\exp\left(\frac{\mu_B}{\kBT}\right).
		\end{align}
	\end{subequations}
	Which is the form we use in the main text.
	
	\section{Transition state theory}
	To derive the dynamics in the system we need to calculate the reaction flux for each reaction.
	But the detailed balance condition used in the main text restricts the ratio of forward $\sF$ and backward $\sB$ only.
	Here we apply transition state theory (TST) to get an explicit form for $\sF$ and $\sB$.
	As an example we consider the reaction $A\rightleftharpoons B$. TST now states that $A$ and $B$ represent local minima in the free energy landscape and the transition from $A$ to be $B$ happens via a saddle point $T$, which forms an energy barrier between $A$ and $B$.
	Assuming that the transition state has a constant energy $E^\dagger$, the height of forward and backward barrier are $E^\dagger-E_A$ and $E^\dagger-E_B$ respectively.
	If the molecule is close to thermal equilibrium the transition is thermally activated and the probability to cross the barrier is given by the height of the barrier times the attempt frequency $\nu$ given by the vibrational modes of the molecule.
	In this case the forward and backward rate of a single molecule are given by
	\begin{subequations}
		\begin{align}
		k^\mathrm{f}=\nu_1\exp\left(-\frac{E^\dagger-E_A}{\kBT}\right),\\
		k^\mathrm{b}=\nu_1\exp\left(-\frac{E^\dagger-E_B}{\kBT}\right).
		\end{align}
	\end{subequations}
	Assuming $\nu$ is the same for both states. Note that $E_i = \mu_i - \kBT \ln(\phi_i)$. Each $A(B)$ particle has a probability of $k^{\mathrm{f}}(k^{\mathrm{b}})$ per unit time to become a $B(A)$ particle. In this case the total forward, $\sF$, and backward, $\sB$, rates are given by the rate per particle, $k^{\mathrm{f}/\mathrm{b}}$, multiplied by the number of particles, given by the local volume fraction: $\sF=\phi_Ak^\mathrm{f}$ and $\sB=\phi_B k^\mathrm{b}$, which results in
	\begin{subequations}
		\begin{align}
		\sF= \nu_1 \exp\left(-\frac{E^\dagger-\mu_A}{\kBT}\right),\\
		\sB = \nu_1 \exp\left(-\frac{E^\dagger-\mu_B}{\kBT}\right),
		\end{align}
	\end{subequations}
	which automatically fulfills the detailed balance condition.
	Applied to the two reactions discussed in the main text, we end up with
	\begin{subequations}
		\begin{align}
		s_1 &= \nu_1 \exp\left(-\frac{E_1^\dagger}{\kBT}\right) \left(\exp\left(\frac{\mu_A}{\kBT}\right) - \exp\left(\frac{\mu_B}{\kBT}\right)\right) = \alpha_1 \left(\exp\left(\frac{\mu_A}{\kBT}\right) - \exp\left(\frac{\mu_B}{\kBT}\right)\right),\\
		s_2 &= \nu_2 \exp\left(-\frac{E_2^\dagger}{\kBT}\right) \left(\exp\left(\frac{\mu_A+\mu_W}{\kBT}\right) - \exp\left(\frac{\mu_B+\mu_F}{\kBT}\right)\right) = \alpha_2 \exp\left(\frac{\mu_F}{\kBT}\right)\left(\exp\left(\frac{\mu_A-\Delta\mu}{\kBT}\right) - \exp\left(\frac{\mu_B}{\kBT}\right)\right).
		\end{align}
	\end{subequations}
	In this case, the parameter $\eta=\exp(\mu_F/\kBT)\alpha_2/\alpha_1$ becomes a constant as $\mu_B$ cancels and we end up as
	\begin{align}
	\eta = \frac{\alpha_2\exp\left(\frac{\mu_F}{k_BT}\right)}{\alpha_1}.
	\end{align}
	So $\eta$ depends on: $\mu_F$, $\mu_W$, $E^\dagger_{1/2}$ and $\nu_{1,2}$. So the total reaction flux ends up as
	\begin{align}\label{Sieq:Rflux}
	\frac{s}{\alpha_1} = \left(\exp\left(\frac{\mu_A}{k_BT}\right) - \exp\left(\frac{\mu_B}{k_BT}\right)\right) + \eta \left(\exp\left(\frac{\mu_A-\Delta\mu}{k_BT}\right) - \exp\left(\frac{\mu_B}{k_BT}\right)\right)
	\end{align}
	
	\section{Catalyzed reactions}
	In a catalyzed reaction the equilibrium is unaffacted by the catalyst while the reaction rates are sped up significantly (generally in both directions).
	Here we use the simple reaction $A+E\rightleftharpoons E_\dagger \rightleftharpoons B+E$, where the catalyst $E$ speeds up the transition between $A$ and $B$, to describe the transition state theory with catalysts.
	The total reaction flux is given by $s=\sF - \sB$, where $\sF=\phi_A\phi_Ek^{f}$ and $\sB = \phi_B\phi_Ek^b$ and introducing the transition state $E_\dagger$ we get for the individual rates
	\begin{align}
	k^f = \alpha \exp\left(-\frac{E_{E_\dagger}-E_A-E_E}{\kBT}\right),\\
	k^b = \alpha \exp\left(-\frac{E_{E_\dagger}-E_B-E_E}{\kBT}\right).
	\end{align}
	Here $E_i=w_i + h_i$ contains the internal energy $w_i$ and the enthalpic interactions with the environment $h_i$ of species $i$. If we assume that the enzyme itself does not change significantly during the reaction, we approximate $h_{E_\dagger}\approx h_E + (h_A+h_B)\delta /2$, where $\delta$ describes how much the enthalpy of the protein deviates from the average between state $A$ and $B$ in the transition state.
	With this approximation we get for the reactive fluxes
	\begin{equation}
	s=\alpha \phi_E\exp\left(-\frac{w_{E_\dagger}-w_E+(h_A+h_B)\delta/2}{\kBT}\right)\left[\exp\left(\frac{\mu_A}{\kBT}\right)-\exp\left(\frac{\mu_B}{\kBT}\right)\right].
	\end{equation}
	Now in general the value of $(h_A+h_B)\delta/2$ can be different inside and outside the droplet, while the enzyme volume fraction is always much larger in the droplet phase compared to the solvent phase $\phi_E^{\ins}\gg\phi_E^\outs$. Oftentimes the role of an enzyme, which catalyzes biochemical reactions, is to bind to the reactant/product molecules and force them into a conformation that is favorable for the reaction (the transition state). In this case the large enzyme 'absorbs' the molecules and the interaction between the molecules and the surrounding becomes neglibible. Therefore we neglect the factor $(h_A + h_B)\delta/2$ here and assume the enzymatic reaction rate is constant and limited by the amount of enzyme $\phi_E$. If we furthermore assume that the driven reaction is sped up by the enzyme, e.g. in phosphorylation reactions, we end up with
	\begin{align}
	s = \alpha_E\phi_E\exp\left(\frac{\mu_F}{\kBT}\right)\left[\exp\left(\frac{\mu_A-\Delta\mu}{\kBT}\right)-\exp\left(\frac{\mu_B}{\kBT}\right)\right],
	\end{align}
	which is the form we use in the main text.
	
	\section{Steady states with driven reactions}
	Here we discuss the steady states for the reaction flux in Eq.\ref{Sieq:Rflux} derived in the last section. In this case the total reaction flux $s=s_1+s_2$ and the fluxes of the individual reactions are
	\begin{subequations}
		\begin{align}
		s_1 &=\alpha_1 \left[\exp\left(\frac{\mu_A}{\kBT}\right) - \exp\left(\frac{\mu_B}{\kBT}\right)\right] ,\\
		s_2 &=\alpha_2 \exp\left(\frac{\mu_F}{\kBT}\right) \left[\exp\left(\frac{\mu_A-\Delta\mu}{\kBT}\right) - \exp\left(\frac{\mu_B}{\kBT}\right)\right].
		\end{align}
	\end{subequations}
	Using $\eta=\alpha_2\alpha_1^{-1}\exp(\mu_F/\kBT)$, $\mu_+=\mu_A+\mu_B$ and $\mu_-=\mu_A-\mu_B$ we write the total reaction flux as
	\begin{equation}
	s(\mu_+,\mu_-) = \alpha_1\exp\left(\frac{\mu_+}{2\kBT}\right)\left[(1+\eta e^{-\Delta\mu/\kBT})\exp\left(\frac{\mu_-}{2\kBT}\right) -(1+\eta)\exp\left(-\frac{\mu_-}{2\kBT}\right)\right].
	\end{equation}
	The total flux vanishes if the term in the square brackets vanishes. This is the case if $\mu_-=\mu_-^*=\mu_A^*-\mu_B^*=\Delta\mu - \kBT\ln\left[\frac{\exp(\Delta\mu/\kBT)+\eta}{1+\eta}\right]$. Note that $\mu_-=0$ corresponds to the equilibrium of the first reaction and $\mu_-=\Delta\mu$ is the equilibrium chemical potential of the driven reaction. Consequently $\mu_-^*$ lies somewhere between $0$ and $\Delta\mu$ and is determined by the ratio of the two rates $\eta$.\\
	Now we make use of $\mu_-^*$ to write the total flux as
	\begin{equation}\label{Sieq:Rfluxs}
	s(\mu_+,\mu_-) = 2\alpha_1\left(1+\eta \exp\left(\frac{-\Delta\mu}{\kBT}\right)\right)\exp\left(\frac{\mu_++\mu_-^*}{2\kBT}\right)\sinh\left[\frac{\mu_--\mu_-^*}{2\kBT}\right],
	\end{equation}
	which shows that $s=0$ is equivalent to $\mu_-=\mu_-^*$. Note that the prefactor depends on the value of the sum of chemical potentials $\mu_A+\mu_B$. $\mu_A+\mu_B$ has to be constant in the whole system close to the steady state and depends on the total amount of protein and the value of $\mu_-^*$.
	
	\section{Uniqueness of the steady state solution}
	First consider the steady state solutions for the sum and difference of the continuity equations for $A$ and $B$, $\partial_t(\phi_A+\phi_B)=0$ and $\partial_t(\phi_A-\phi_B)=0$ (for simplicity $\Lambda_A=\Lambda_B$):
	\begin{align}
	\nabla^2(\mu_A+\mu_B) = 0,\\
	\nabla^2(\mu_A-\mu_B) - 2s=0.
	\end{align}
	Where we assume no flux boundary conditions ($\bf{n}\nabla\mu_i=0$ at the boundary). Introducing the notation $\mu_+=\mu_A+\mu_B$ and $\mu_-=\mu_A-\mu_B$, we see that $\mu_+=\mathrm{const}$ everywhere.
	Solving the second equation requires the knowledge of $s$. In the simplest case we can linearize the total reaction flux derived in the previous section, so $s(\mu_-) \approx \alpha \mu_-$. In this case the equation
	\begin{equation}
	\nabla^2\mu_- - 2\alpha\mu_-=0
	\end{equation}
	has the trivial solution $\mu_-=0$. To show that the solution is unique we assume we had two solutions $\mu_1$ and $\mu_2$. Because the equation is linear, $f=\mu_1-\mu_2$ is a solution as well.
	Now we use that
	\begin{equation}
	\nabla(f\nabla f) = (\nabla f)^2 + f\nabla^2f = (\nabla f)^2 + 2\alpha f^2.
	\end{equation}
	Integrating this over the whole volume results in
	\begin{equation}
	\int_V \nabla(f\nabla f)\,dV = \int_V (\nabla f)^2 + 2\alpha f^2\,dV.
	\end{equation}
	Applying Gauß integral theorem we arrive at
	\begin{equation}
	\oint_{\partial V} f\nabla f\,d\mathbf{S} = \int_V (\nabla f)^2 + 2\alpha f^2\,dV.
	\end{equation}
	Now on the boundary the outward normal of $\nabla f$ vanishes and thus the boundary integral over $f\nabla f$ vanishes. But $f^2$, $2\alpha$ and $(\nabla f)^2$ are all non-negative, so that $f$ and $\nabla f$ have to vanish everywhere and thus $\mu_1=\mu_2$.\\
	How about the nonlinear version where $s$ is given by Eq.\ref{Sieq:Rfluxs}? First $\mu_+=\mathrm{const}$ remains necessarily fulfilled, but we need to find the solution for
	\begin{equation}
	\nabla^2\bar{\mu}_- - a^2\sinh\left[\bar{\mu}_-\right]=0,
	\end{equation}
	with $\bar{\mu}_-=\frac{\mu_--\mu_-^*}{2\kBT}$ now. We can repeat the same steps as before, but with $\bar{\mu}_-$ instead of $f$ and end up with
	\begin{equation}
	\oint_{\partial V} \bar{\mu}_-\nabla \bar{\mu}_-\,d\mathbf{S} = \int_V (\nabla \bar{\mu}_-)^2 + 2\alpha \bar{\mu}_-\sinh[\bar{\mu}_-]\,dV.
	\end{equation}
	Now $\sinh(\bar{\mu}_-)$ has the same sign as $\bar{\mu}_-$, so $\sinh(\bar{\mu}_-)\bar{\mu}_-\geq 0$ and thus the only solution is $\bar{\mu}_-=0$ or $\mu_- = \mu_-^*$.

	\section{Reaction rate in droplet and solvent phase}
	In this section we discuss the approximation for the reaction rate $s$ in droplet and solvent phase with enzymes segregating in the droplet phase. The general form for $s$ as a function of $\mu_+$ and $\mu_-$ is (Eq.\ref{Sieq:Rfluxs}):
	\begin{equation}
	s(\mu_+,\mu_-)=2\alpha_1(1 + e^{-\frac{\Delta\mu}{\kBT}}\eta) e^{\frac{\mu_++\mu_-^*}{2\kBT}}\sinh\left[\frac{\mu_--\mu_-^*}{2\kBT}\right].
	\end{equation}
	For the reaction flux outside, we assume that the distance between droplets is large compared to the reaction diffusion length scale. In this case $s^\outs=0$ far away from the droplet interface and consequently $\mu^\outs_- = \mu_-^{*,\outs}$. 
	Expanding $\sinh(x)\approx x + \mathcal{O}(x^3)$, we get that
	\begin{equation}
	s^\outs\approx \alpha_1(1 + e^{-\frac{\Delta\mu}{\kBT}}\eta) e^{\frac{\mu_++\mu_-^*}{2\kBT}}\left[\frac{\mu_--\mu_-^*}{\kBT}\right]
	\end{equation}
	This can be further approximated for strong driving, where $e^{-\frac{\Delta\mu}{\kBT}}\eta\ll 1$ and thus $\mu_-^{*,\outs}\approx \kBT\ln(1+\eta)$:
	\begin{equation}
	s^\outs\approx \alpha_1\sqrt{(1+\eta) e^{\frac{\mu_+}{\kBT}}}\left[\frac{\mu_-}{\kBT}-\ln(1+\eta)\right]
	\end{equation}
	
	In the droplet phase the driven reaction is sped up by a factor $\Gamma\gg 1$ due to segregation of enzymes. Here we consider the case where $\mu_-^{*,\ins}\gg\mu_-(R)>\mu_-^{*,\outs}$. In this case we assume that the variation of $\mu_-^\ins(r)$ is small compared to the difference between $\mu_-^{*,\ins}-\mu_-(R)$ (or that the radius $R$ of the droplet is small compared to the reaction diffusion length scale). We then expand $s
	^\ins$ around $\mu_-(R)$ and get
	\begin{equation}
	s^\ins\approx \alpha_1(1 + e^{-\frac{\Delta\mu}{\kBT}}\Gamma\eta) e^{\frac{\mu_++\mu_-^{*,\ins}}{2\kBT}}e^{\frac{\mu_-^{*,\ins}-{\mu_-(R)}}{2\kBT}}\left(\frac{\mu_-^\ins - \mu_-(R)}{2\kBT}-1\right),
	\end{equation}
	Which we write as
	\begin{equation}
	s^\ins\approx \alpha_1(1 +\Gamma\eta) e^{\frac{\mu_+-\mu_-(R)}{2\kBT}}\left(  \frac{\mu_-^\ins - \mu_-(R)}{2\kBT} - 1\right).
	\end{equation}
	Now we assume strong segragation such that $\Gamma\eta\gg1$ and we end up with
	\begin{equation}\label{Eq:ReactFlIn}
	s^\ins \approx \Gamma\eta \alpha_1e^{\frac{\mu_+-\mu_-(R)}{2\kBT}}\left(  \frac{\mu_-^\ins - \mu_-(R)}{2\kBT} - 1\right).
	\end{equation}
	Note that $\frac{\mu_+-\mu_-(R)}{2\kBT}=-\frac{\mu_B^{eq}}{\kBT}$ and for the reaction outside
	\begin{equation}\label{Eq:ReacFlOut}
	s^\outs\approx \alpha_1(1+\eta)^{1/2} e^{\frac{\mu_+}{2\kBT}}\left[\frac{\mu_-}{\kBT}-\ln(1+\eta)\right].
	\end{equation}
	Thereby the reaction flux in both phases takes the form
	\begin{equation}
	s^{\ins/\outs} = \alpha_{\ins/\outs} (\mu_-^{\ins/\outs}-\mu_{-,0}^{\ins/\outs}),
	\end{equation}
	where $\alpha_{\ins/\outs}$ is the rate and $\mu_{-,0}^{\ins/\outs}$ is the chemical potential for which $s^{\ins/\outs}$ vanishes. Written out, we get $\alpha_\outs = \alpha_1(1+\eta)^{1/2}e^{\frac{\mu_+}{2\kBT}}/\kBT$ and $\mu_{-,0}^\outs=\kBT\ln(1+\eta)$ and $\alpha_\ins = \Gamma\eta\alpha_1e^{\frac{\mu_+-\mu_-(R)}{2\kBT}}/(2\kBT)$ and $\mu_{-,0}^\ins = \mu_-(R) + 2\kBT$.

	\section{Constructing a non-equilibrium steady state solution}
	We consider a steady state of a single droplet in a solvent phase. Then the steady state continuity equation for droplet and solvent phase are
	\begin{subequations}
		\begin{align}
		\Lambda\nabla^2\mu_+^\ins &= 0\\
		\Lambda\nabla^2\mu_-^\ins &= 2 s^\ins
		\\
		\Lambda\nabla^2\mu_+^\outs &= 0\\
		\Lambda\nabla^2\mu_-^\outs &= 2 s^\outs.
		\end{align}
	\end{subequations}
	We furthermore assume the system to be radially symmetric so that $\mu_{+,-}(r)$ and we have no flux boundary conditions at $r=0,L$ where $L$ is the radius of the system.
	Both $\mu_-$ and $\mu_+$ have to be continous across the interface, which together with $\mu_+^\ins=const$ and $\mu_+^\outs=const$ implies that $\mu_+$ is constant in the whole system.
	Furthermore we know that at the droplet interface $\mu_B$ is fixed by the maxwell construction, thus $\mu_B^\ins(R)=\mu_B^\outs(R)=\mu_B^{eq}$ and thus $\mu_-(R) = \mu_-^R=\mu_+(R) - 2 \mu_B^{eq}$.
	To get a solution for $\mu_-(r)$, we need the reactive flux $s$ as a function of $\mu_-$. Importantly due to different enzyme content, and thus $\mu_-^*$, in droplet and solvent phase, $s^\ins$ and $s^\outs$ are different.\\
	We derived the approximate reaction fluxes inside and outside above, see Eq.\ref{Eq:ReactFlIn} and \ref{Eq:ReacFlOut}
	\begin{equation}\label{Eq:LinReactOut}
	2s^\outs\approx 2\alpha_1\sqrt{(1+\eta) e^{\frac{\mu_+}{\kBT}}}\left[\frac{\mu_--\kBT\ln(1+\eta)}{\kBT}\right]=\alpha_\outs \left(\mu_--\kBT\ln(1+\eta)\right).
	\end{equation}
	Where the rate $\alpha_{\outs}$ has units per time per $\kBT$. In this case the solution for the chemical potential in the solvent phase is given by
	\begin{equation}
	\mu_-^\outs(r) = \kBT\ln(1+\eta) + (\mu_-^R - \kBT\ln(1+\eta))\frac{R}{r}e^{-(r-R)/\ell_\outs},
	\end{equation}
	where $\ell_\outs^2=\Lambda/\alpha_\outs$ is the (squared) reaction diffusion length scale.
	We can calculate the total reaction flux outside the droplet $S^\outs$ using Eq.\ref{Eq:LinReactOut} and the radial profile of $\mu_-^\outs$
	\begin{equation}
	S^\outs = 4\pi\int_R^\infty \alpha_\outs^2(\mu_-^\outs(r) - \mu_{-,0}^{\outs})r^2 \,dr = 4\pi\alpha^2_\outs (\mu_-^R - \kBT\ln(1+\eta))\int_R^\infty r Re^{-(r-R)/\ell_\outs} \,dr.
	\end{equation}
	In the (quasi) steady state\cite{Weber_RepProgPhys_2019}, the total reaction flux has to be equal to the diffusive flux into/out of the droplet $J=S^\outs$. And $J$ corresponds to the supply ($J>0$) or depletion ($J<0$) of droplet material and thus determines droplet growth.\\
	Plugging in $\mu_-^\outs(r)$, we get
	\begin{equation}\label{Sieq:Diffflux}
	J = 4\pi R\Lambda(\mu_-^R-\kBT\ln(1+\eta))\left(\frac{R}{\ell_\outs} + 1 \right).
	\end{equation}
	So the total diffusive flux into the droplet scales with $R$ for large reaction diffusion length $\ell^\outs$ and with $R^2$ for small $\ell^\outs$.\\
	Now in the droplet phase, the driven reaction is approximately sped up by a factor of $\Gamma\gg1$ due to the presence of enzymes. And we derived for the reaction flux in the droplet (see Eq.\ref{Eq:ReactFlIn})
	\begin{align}
	2s^\ins \approx 2\Gamma\eta \alpha_1e^{\frac{\mu_+-\mu_-^R}{2\kBT}}\left(  \frac{\mu_-^\ins - \mu_-^R}{2\kBT} - 1\right) = -\alpha_\ins\Gamma\left(2\kBT+\mu_-^R-\mu_-\right).
	\end{align}
	Solving the steady state equation inside the droplet, we get the chemical potential profile inside the droplet which reads
	\begin{equation}
	\mu_-^\ins(r) = \mu_-^R + 2\kBT \left(1- \frac{R}{r}\frac{\sinh(r/\ell_\ins)}{\sinh(R/\ell_\ins)}\right),
	\end{equation}
	where $\ell_\ins^2 = \Lambda/(\alpha_\ins\Gamma)$ is the squared reaction diffusion length scale inside the droplet. Fro droplets small compared to the reaction diffusion length scale (which is necessary for droplet size control) $R/\ell_\ins\ll 1$ and $r<R$, we get:
	\begin{equation}
	\mu_-^\ins \approx \mu_-^R + \frac{2\kBT}{3\ell_\ins^2}(R^2-r^2) = \mu_-^R + \frac{2\kBT \alpha_\ins\Gamma}{3\Lambda}(R^2-r^2).
	\end{equation}
	we can then calculate the total reaction flux in droplet phase, $S^\ins = 4\pi\int_0^R r^2 s^\ins dr$:
	\begin{equation}
	S^\ins = -4\pi (\alpha_\ins\Gamma) (2\kBT)\int_0^R r^2\left(1-\frac{s_0}{3\Lambda}(R^2-r^2)\right)dr = -(\alpha_\ins\Gamma)(2\kBT)V_R\left(1-\frac{2R^2}{15\ell_\ins^2}\right).
	\end{equation}
	For the calculation of the stable droplet radius we assume that $R^2\ll \ell_\ins^2$ and call $S^\ins=S$, suhc that
	\begin{align}
	S\approx -(\alpha_\ins\Gamma)(2\kBT)V_R.
	\end{align}
	The stable droplet radius is then determined by the condition $S+J=0$, which is given by either 
	\begin{equation}
	-(\alpha_\ins\Gamma)V_R(2\kBT) + 4\pi R\Lambda(\mu_-^R-\kBT\ln(1+\eta)) = 0
	\end{equation}
	and thus
	\begin{equation}\label{Sieq:Rstab1}
	R_*^2 = \frac{3\Lambda}{2\alpha_\ins\Gamma} (\mu_-^R-\kBT\ln(1+\eta))= \frac{3\ell_\ins^2}{2}\left(\frac{\mu_-^R}{\kBT}-\ln(1+\eta)\right)\propto \exp(-\chi_E),
	\end{equation}
	for $\ell_\outs\gg R$ or for $\ell_\outs\ll R$
	\begin{align}
	-(\alpha_\ins\Gamma)V_R(2\kBT) + \frac{4\pi\Lambda R^2}{\ell_\outs}(\mu_-^R-\kBT\ln(1+\eta))=0,
	\end{align}
	which corresponds to
	\begin{align}\label{Sieq:Rstab2}
	R_* = \frac{3\ell_\ins^2}{2\ell_\outs}\left(\frac{\mu_-^R}{\kBT}-\ln(1+\eta)\right)\propto \exp(-\chi_E).
	\end{align}
	So depending on the reaction diffusion length scale in the dilute phase, the stable droplet radius scales with $R_*\propto\exp(-\chi_E/2)$ ($\ell_\outs\gg R$) or $R_*\propto\exp(-\chi_E)$ ($\ell_\outs\ll R$).

	\section{Entropy production rate}
	An important quantity to describe non-equilibrium states is the total entropy production rate $\sigma$, which vanishes in equilibrium systems, but is always positive in a non-equilibrium state.
	The entropy production rate measures the local entropy production per unit time. Thereby each thermodynamic process (in our case the two reactions and diffusion) has its own entropy production rate given by the product of the thermodynamic force (here chemical potential differences) and flux (here diffusive flux and reaction rates).
	For our system the local entropy production rate is given by
	\begin{equation}
	\sigma = \sum_i -\textbf{j}_i\nabla\mu_i + s_1(\mu_A-\mu_B) + s_2(\mu_A-\mu_B-\Delta\mu).
	\end{equation}
	We are interested in the entropy production rate in the non-equilibrium steady state to qunatify the energy cost of the given steady state.
	In this case all forces and fluxes, and thus the entropy production rate,  are time independent.
	In a first step we discuss the non-equilibrium steady state without segragation of enzymes, where $\mu_-(x,t)=\mu_-^*$, $s_1=-s_2$ and $\nabla\mu_i=0$. There the entropy production is just given by\cite{DeGroot_NEQTD_1984,Weber_RepProgPhys_2019}
	\begin{equation}
	\sigma = -s_2(\mu_-^*)\Delta\mu,
	\end{equation}
	where $s_2(\mu_-^*<0$ and $\Delta\mu>0$ so that $\sigma>0$ as required. Thereby the steady state flux $s_2(\mu_-^*)$ is given by
	\begin{equation}
	s_2(\mu_-^*) = -\alpha_2e^{(\mu_B+\mu_F)/\kBT}\left(\frac{1-e^{-\Delta\mu/\kBT}}{1+\eta e^{-\Delta\mu/\kBT}}\right).
	\end{equation}
	In the case of segregating enzymes, the chemical potential gradients do not vanish anymore and the entropy production is not only determined by the reactions anymore. In the steady state the total entropy production is equal to the external energy input, such that
	\begin{align}
	\sigma_* = -\Delta\mu s_2(\mu_-(r),\mu_+),
	\end{align}
	but now $\mu_-(r)$ is a more complex function of the radius.
	
	\section{Estimating the steady state radius}
	Here we try to estimate conditions under which droplet size control is possible in cells and the corresponding droplet sizes. We estimate the diffusivity $D$ for proteins in cells to be approximately $\unit[0.1-10]{\mu m^2/s}$\cite{Schavemaker_2018}.
	
	Estimating reaction rates and comparing them to our reactions is more difficult, because usually mass action kinetics are used to describe reactions. In addition enzymatic reactions are often described using Michaelis Menten kinetics. Assuming that the substrate (here the droplet material) is abundant, which is the case in the droplet phase, the rate of the enzymatic michaelis menten reaction can be approximated by a single rate $k_{cat}\sim \unit[0.01-100]{s^{-1}}$\cite{Milo_CellBio_2015}. If the driven reaction dominates in the droplet phase, we can write the total reaction flux as
	\begin{align}
	S\approx -k_{cat}\phi_E^\ins V_R.
	\end{align}
	To get the diffusive flux $J$ into the droplet, we make use of the fact that $\mu_+$ is constant to write $\mu_-=\mu_+-2\mu_B$ and thus $\mu_-^R-\kBT\ln(1+\eta)=2(\mu_{B,*}-\mu_B^R)$. Furthermore we know that $\phi_B$ is fixed by $\phi_B^\outs$ at the interface and $\phi_B\approx K/(1+K)\bar{\phi}_+$ far away from the droplet. The second condition is valid if the reaction flux vanishes far away from the droplet and the total droplet volume is small. If the protein and enzyme are dilute in the solvent phase $\phi_C
	^\outs\approx 1$ and the concentration variations are small, we can linearize the chemical potential according to $\mu_B\approx \mu_B(\phi_B^{eq}) + \kBT\frac{\phi_B-\phi_B^{eq}}{\phi_B^{eq}}$.\\
	In total the diffusive flux into the droplet is then given by (see Eq.\ref{Sieq:Diffflux}
	\begin{align}
	J \approx 4\pi R\frac{\Lambda \kBT}{\phi_B^{eq}}\left(\frac{K}{1+K}\bar{\phi}_+ - \phi_B^{eq}\right)=4\pi R D\left(\frac{K}{1+K}\bar{\phi}_+ - \phi_B^{eq}\right),
	\end{align}
	With $D=\Lambda\kBT/\phi_B^{eq}$.
	
	In this case the stable radius, determined by $S+J=0$ is given by
	\begin{align}
	-k_{cat}\phi_E^\ins V_R + 4\pi R D \left(\phi_B^{eq}-\frac{K}{1+K}\bar{\phi}_+\right)=0
	\end{align}
	with $K\approx (1+\eta)^{-1}\exp(w_A-w_B)$ and we write $D=\Lambda (\partial_{\phi_B}\mu_B)|_{\phi_B^{eq}}$.\\
	In this case, the stable droplet radius, for small reaction diffusion length in the dilute phase, is given by
	\begin{subequations}
		\begin{align}
		R_*^2 = \frac{3D}{k_{cat}}\frac{\Delta\phi_B}{\phi_E^\ins},\\
		R_*^2 = \frac{3D}{k_{cat}}\frac{v_E\Delta c_B}{v_Bc_E^\ins}
		\end{align}
	\end{subequations}
	where the second form, expressed in terms of particle concentration instead of volume fraction, is the one used in the main text. For the rates and diffusivities mentioned above, $k_{cat}\sim\unit[0.01-100]{s^{-1}}$ and $D\sim \unit[0.1-10]{\mu m^2/s}$ and approximating the ration of droplet protein concentration difference to enzyme concentration $\Delta c_B/c_E^\ins\sim 0.1-10$ for $v_E\approx v_B$, we get stable droplet radii on the order of $R_*\sim \unit[10
	^2-10^{-2}]{\mu m}$.

\end{document}